\documentclass[a4paper,fleqn,usenatbib]{mnras}
\usepackage[T1]{fontenc}
\usepackage{ae,aecompl}
\usepackage{amsmath}
\usepackage{amsfonts}
\usepackage{amssymb}
\usepackage{graphicx}
\usepackage{color,xcolor}
\def\mnras{MNRAS}
\def\aap{A$\&$A}
\def\nat{Nature}
\def\apj{ApJ}
\def\apjl{ApJ Letters}
\def\apjs{ApJS}

\def\aj{AJ}

\newcommand{\msun}{\, {\rm M}_{\odot}}
\newcommand{\msunyr}{\, {\rm M}_{\odot}\,{\rm yr^{-1}}}
\newcommand{\rsun}{\, {\rm R}_{\odot}}
\newcommand{\au}{\, {\rm au}}

\title[Neverland's Parameters]{Peter Pan Discs: finding Neverland's parameters}
\author[G. A. L. Coleman and T. J. Haworth]{Gavin A. L. Coleman\thanks{Email: gavin.coleman@qmul.ac.uk}, Thomas J. Haworth\\
Astronomy Unit, School of Physics and Astronomy, Queen Mary University of London, Mile End Road, London, United Kingdom}
\date{Accepted 2020 May 18. Received 2020 May 18; in original form 2020 April 21}
\pubyear{2020}
\begin{document}
\label{firstpage}
\pagerange{\pageref{firstpage}--\pageref{lastpage}}
\maketitle
\begin{abstract}
Peter Pan discs are a recently discovered class of long-lived discs around low-mass stars that survive for an order of magnitude longer than typical discs. In this paper we use disc evolutionary models to determine the required balance between initial conditions and the magnitude of dispersal processes for Peter Pan discs to be primordial. We find that we require low transport ($\alpha\sim10^{-4}$), extremely low external photoevaporation ($\leq10^{-9}\msunyr$), and relatively high disc masses ($>0.25M_*$) to produce discs with ages and accretion rates consistent with Peter Pan discs. Higher transport ($\alpha = 10^{-3}$) results in disc lifetimes that are too short and even lower transport($\alpha = 10^{-5}$) leads to accretion rates smaller than those observed. The required external photoevaporation rates are so low that primordial Peter Pan discs will have formed in rare environments on the periphery of low-mass star-forming regions, or deeply embedded, and as such have never subsequently been exposed to higher amounts of UV radiation. Given that such an external photoevaporation scenario is rare, the required disc parameters and accretion properties may reflect the initial conditions and accretion rates of a much larger fraction of the discs around low-mass stars.
\end{abstract}

\section{Introduction}
\label{sec:intro}
With the discovery of over 4000 very diverse exoplanets \citep[e.g.][]{2015ARA&A..53..409W}, understanding the formation of planets from circumstellar discs of material is one of the key problems in astrophysics. Huge advances have been made in recent years, with direct probes of the mid-plane structure of discs \citep[e.g.][]{2015ApJ...808L...3A, 2018ApJ...869L..41A} and more statistical assessments of the properties of discs in different stellar clusters \citep[e.g.][]{2017AJ....153..240A, 2018ApJ...860...77E}. A well-established result from the latter class of study is the typical lifetime of discs, which is derived by inferring the fraction of stars with circumstellar discs in clusters of different ages \citep[e.g.][]{Mamajek09, 2014A&A...561A..54R}. This shows that the typical lifetime is $<3\,$Myr and that essentially all discs are dispersed by 10\,Myr. It has also been shown that these lifetimes are a function of the stellar mass, with discs around lower mass living longer \citep{Carpenter06}. However nearly all of the discs around low-mass stars are still found to have dispersed within 20 Myr \citep{PecautMamajek16}.

In light of the above results, a new class of long-lived discs around low-mass stars have recently been discovered, termed ``Peter Pan'' discs \citep{LeeSongMurphy20,Silverberg20}.
For example \citet{Silverberg16} observed that the 45-Myr-old $0.11\msun$ M-dwarf WISE J080822.18-644357.3 has an infrared excess (indicative of circumstellar material) and in the same system \citet{Murphy18} found significant $H_{\alpha}$ emission (indicative of accretion).
\citet{Flaherty19} observed this system with ALMA, finding a $<16\au$ mm continuum disc but no CO gas detection.
The lack of CO detection, and hence possibly gas, is apparently at odds with the measured accretion. However, there is evidence that primordial CO depletes on approximately few Myr time-scales \citep[e.g.][]{Zhang19}, while secondary generations of CO due to processes like collisions \textit{are} observable in debris discs \citep[e.g.][]{Matra17}. It may therefore be the case that the disc is simply CO dark.
A number of other Peter Pan discs have now been discovered with ages ranging from 42 to 55\,Myr, all around low-mass stars \citep{Silverberg20}.
Although the frequency with which these Peter Pan discs occur is currently unknown, it is important for understanding disc evolution and planet formation generally to understand why these discs can survive for more than an order of magnitude longer than typical disc lifetimes. Making a first attempt at doing so is the objective of this paper. 

Key to understanding how discs can survive for a long time is the disc dispersal processes. Discs are depleted and dispersed in a number of ways, including the following:
\begin{enumerate}
    \item Gas accretion through the disc onto the star \citep[e.g.][]{Lynden-BellPringle1974,Suzuki09, 2016A&A...591L...3M}.
    \item The host star drives a photoevaporative wind from the inner disc \citep[e.g.][]{2017RSOS....470114E}.
    \item The ambient stellar cluster UV field drives a photoevaporative wind from the outer disc \citep[e.g.][]{2004ApJ...611..360A, 2016MNRAS.457.3593F, 2019MNRAS.485.3895H}.
    \item Gas is accreted by growing planets \citep[e.g.][]{CPN17}.  
\end{enumerate}
How these processes interplay and dominate at different epochs is still an area of active research. 

In addition to dispersal, the initial conditions of discs are still only weakly constrained due to the rapid and highly obscured nature of disc assembly and early evolution. Although there are some examples of class 0 discs being discovered \citep{2013A&A...560A.103M}, the properties of discs in clusters have only been surveyed down to $\sim1\,$Myr \citep[e.g.][]{2014A&A...561A..54R}. Furthermore, \cite{Haworth20} recently demonstrated that low-mass stars could support axisymmetric discs that are much more massive than previously thought.  

Peter Pan discs are likely to require unusually slow dispersal and/or unusually massive initial conditions. In this paper, we use disc evolutionary models to explore which combinations of the above factors permit the existence of Peter Pan discs. 
 
\vspace{-0.4cm} 
\section{Physical Model and Parameters}
\label{sec:ppds}

\vspace{-0.1cm}
\subsection{Gas disc model}
The gas disc model that we use closely follows that presented in \citet{ColemanNelson14,ColemanNelson16}, with the addition of a photoevaporative wind that is driven by UV radiation from external sources, i.e. nearby massive stars.
To determine the evolution of the gas disc, we solve the standard diffusion equation for a 1D viscous $\alpha$-disc model \citep{Shak}, but stress that the viscous prescription used here is meant only as a proxy for the mass flow through the disc and that other processes that induce a mass flow through the disc \citep[e.g. magnetically driven disc winds;][]{Kunitomo20} should yield similar results.
Disc temperatures are calculated by balancing blackbody cooling against viscous heating and stellar irradiation.
We include an active turbulent region that increases the value of $\alpha$ in the inner regions of disc where $T\geq1000$ K, mimicking the effect of fully developed turbulence forming in these regions \citep{UmebayashiNakano1988,DeschTurner2015}.
While the disc loses mass when it is accreted by the central star, it also loses mass through a photoevaporative wind due to the high-energy photons emanating from the central star.
This wind is driven from the surface layers of the disc \citep{Dullemond} and we follow \citet{ColemanNelson14} for its implementation.\footnote{For low-mass stars, the mass-loss rates through internal photoevaporation are typically $\dot{M}_{\rm int}<10^{-10}\msunyr$. Within our simulations we use different values for the internal photoevaporation rate, but since they are typically much smaller than the mass accretion and external photoevaporation rates, we find negligible differences in disc lifetimes.}

\vspace{-0.2cm}
\subsection{External photoevaporation}
External photoevaporation of a disc by radiation from nearby stars can play an important role in setting the evolution of the  disc mass \citep{2014ApJ...784...82M, 2017AJ....153..240A} radius \citep{2018ApJ...860...77E} and lifetime \citep{2016arXiv160501773G, 2019MNRAS.490.5678C, 2020MNRAS.492.1279S, 2020MNRAS.491..903W} even in weak UV environments \citep{2017MNRAS.468L.108H}.

We follow the Far-UV-driven photoevaporation prescription of \citet{Matsuyama03}, in which the change in the gas surface density $\Sigma$ is equal to
\begin{equation}
\dot{\Sigma}_{\rm w}(R) =  \left\{ \begin{array}{ll}
0, & R\le \beta R_{\rm g, fuv}, \\
\\
\dfrac{\dot{M}_{\rm w}(R_{\rm d})}{\pi(R_{\rm d}^2-\beta^2 R_{\rm g, fuv}^2)},& R>\beta R_{\rm g, fuv}.
\end{array} \right.
\end{equation}
where $R_{\rm g, fuv}$ is the gravitational radius (at which material is unbound) and $\dot{M}_{\rm w}(R_{\rm d})$ is the mass-loss rate for a disc of size $R_{\rm d}$.
We want to explore the effects of different mass-loss rates, so consider the mass-loss rate at the initial disc size $\dot{M}_{\rm w, 0}$ as a free parameter and scale this with the disc radius as it evolves as
\begin{equation}
    \dot{M_w}(R_{\rm d})= \dot{M}_{\rm w, 0} \left(\frac{R_{\rm d}}{R_{\rm ini}}\right),
    \label{equn:evap}
\end{equation}
where $R_{\rm ini}$ is the initial disc size. We do not consider exterme UV (EUV)-driven mass-loss since, as we will demonstrate, Peter Pan discs require low-UV environments ,where EUV radiation is negligible. 

More sophisticated approaches to computing the mass-loss rate do exist, such as the \textsc{fried} grid of mass-loss rates \citep{2018MNRAS.481..452H}. However, as we will show here, Peter Pan discs require \textit{extremely} low external photoevaporation rates that are of the order of floor value in \textsc{fried} and hence limit its applicability. We therefore opt at this stage to retain the simpler but well-controlled approach to setting the external photoevaporation mass-loss rate described above. 

\begin{table}
\centering
\begin{tabular}{lc}
\hline
Parameter & Value\\
\hline
Disc inner boundary & 0.0189 $\au$ (3 d)\\
Disc outer boundary $R_{\rm ini}$ & [50, 100, 200]$\au$\\
Disc mass & [0.2, 0.4, 0.6, 0.8, 1] $\times M_{\rm d, max}$\\
$\alpha$ & [$10^{-3}$, $10^{-4}$, $10^{-5}$]\\
$M_*$ & $0.1 \msun$\\
$R_*$ & $1 \rsun$\\
$T_*$ & 3000 K\\
$\dot{M}_{w, 0}$ & [$10^{-7}$, $10^{-8}$, $10^{-9}$, $10^{-10}$] $\msunyr$ \\
\hline
\end{tabular}
\caption{Disc and stellar model parameters}
\vspace{-0.2cm}
\label{tab:parameters}
\end{table}

\vspace{-0.2cm}
\subsection{Initial disc parameters}

\begin{figure*}
    \centering
    \includegraphics[width=8.5cm]{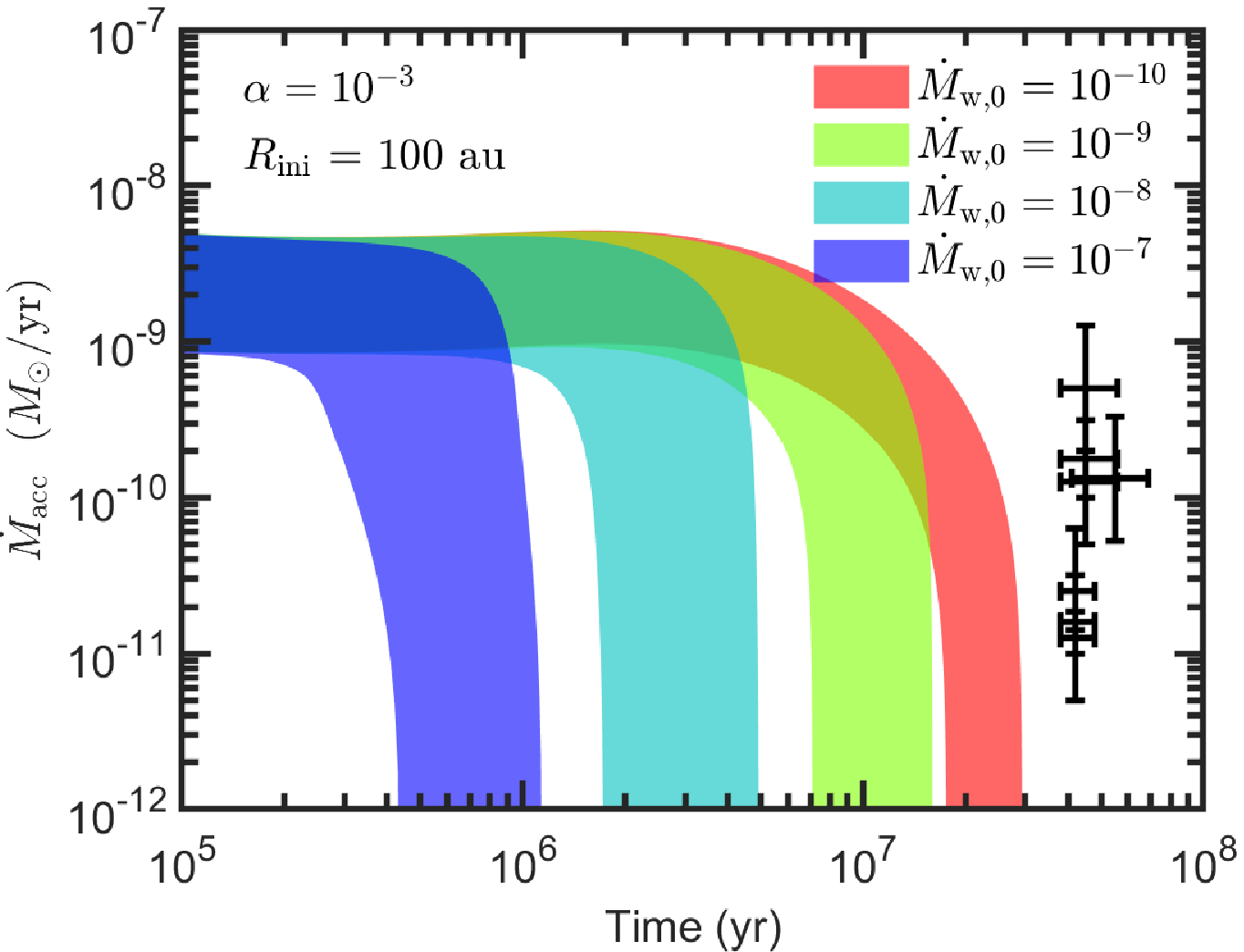}
    \includegraphics[width=8.5cm]{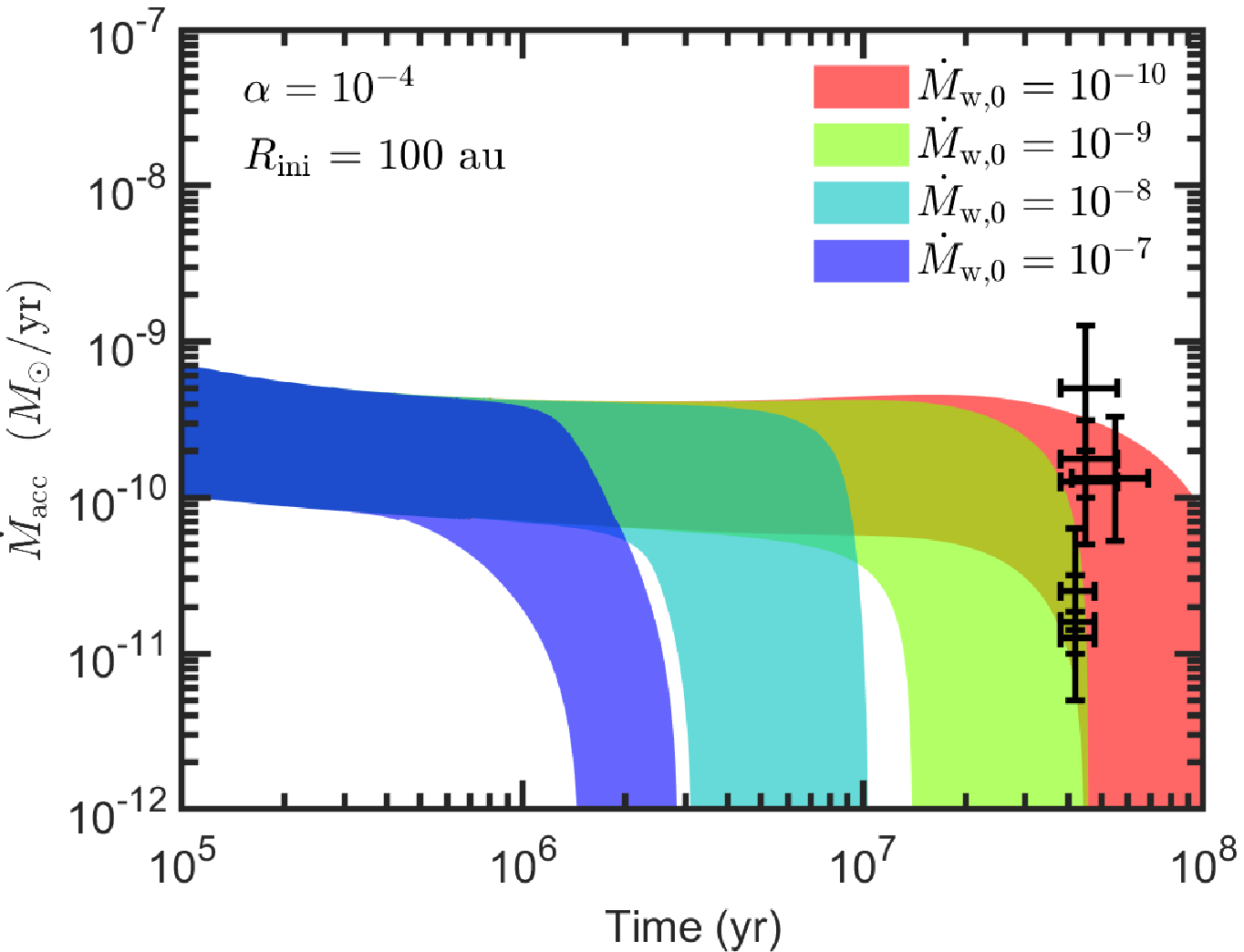}
    \vspace{-0.2cm}
    \caption{Accretion rates over time for our models with viscous $\alpha = 10^{-3}$ ({\it left-hand panel}), and $\alpha = 10^{-4}$ ({\it right-hand panel}). Different colours correspond to different external photoevaporation rates and the thicknesses of the regions are set by the range of initial disc masses. The black points mark the observed accretion rates for Peter Pan discs \citep{LeeSongMurphy20,Silverberg20}. When the external photoevaporation rates are high, the disc lifetimes are too short compared to the Peter Pan discs. However, with lower photoevaporation rates ($\dot{M}_{\rm w,0}\leq10^{-9}\msunyr$), we can match the observed accretion rates for lower $\alpha$ values.}
    \label{fig:AlphaEvap}
\end{figure*}

With the stars in the Peter Pan discs having spectral types of M4--M6 with temperatures consistent with low-mass stars, e.g. $\sim3000$ K \citep{LeeSongMurphy20,Silverberg20}, we run our simulations with a $0.1 \msun$ central star with a radius of $1 \rsun$ and a temperature of $3000$ K.
This is similar to that used in previous work examining the formation of the TRAPPIST-1 planetary system \citep{Coleman19}.
We set the inner edge of the disc to be equal to a period of 3 d, while the outer edge of the disc, $R_{\rm ini}$, is set to 50, 100, or 200 au, to examine the effect it has on the evolution of the disc.
For the initial mass of the disc we follow \citet{Haworth20} where they find the maximum mass that a gas disc of a specific size can be, before it becomes unstable to gravitational instability:
\begin{equation}
    \label{eq:max_disc_mass}
    \dfrac{M_{\rm d, max}}{M_*} < 0.17 \left(\dfrac{R_{\rm ini}}{100\au}\right)^{1/2}\left(\dfrac{M_*}{\msun}\right)^{-1/2}.
\end{equation}
A key point in that work is that the maximum axisymmetric disc-to-star mass ratio increases with decreasing host star mass.
Traditionally disc masses were thought to be $\leq30\%$ of the stellar mass (regardless of stellar mass), and as such with eq. \ref{eq:max_disc_mass} the maximum disc masses for low-mass stars are significantly in excess of that. For discs extending to $200\au$, eq. \ref{eq:max_disc_mass} would result in a maximum disc mass equal to $\sim75\%$ of the stellar mass ($38\%$ for a disc of size $50\au$) while still remaining gravitational stable.

Given the relation for the maximum disc mass, in our simulations we use disc masses of 0.2--1 $\times M_{\rm d, max}$.
To examine the effects of the mass flow through the disc, we use values of $\alpha$ between $10^{-5}$ and $10^{-3}$, while for examining the effects of the mass-loss through external photoevaporation, we use mass-loss rates between $10^{-10}$ and $10^{-7}\msunyr$.
For a full list of our simulations parameters, see Table \ref{tab:parameters}.

\vspace{-0.4cm}
\section{Under what conditions can a Peter Pan disc be produced?}
\label{sec:Results}
Using our grid of evolutionary models, we constrain when Peter Pan discs result depending on the initial disc properties and the magnitude of disc dispersal processes. Ultimately, we find that low accretion and evaporation rates, coupled with high disc masses are required, which we now illustrate in turn. 

\begin{figure}
    \centering
    \includegraphics[width=8.5cm]{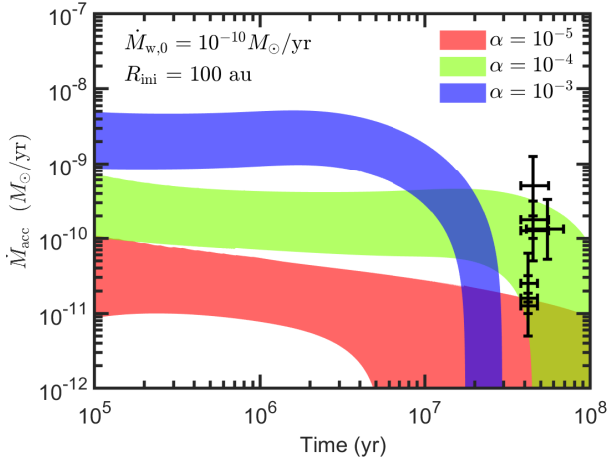}
    \vspace{-0.5cm}
    \caption{Accretion rates over time for our models with a low external photoevaporation rate and various values of viscous $\alpha$ parameter. If $\alpha$ is too high ($\alpha\sim10^{-3}$) the disc lifetime is too short for Peter Pan discs. Conversely, if $\alpha$ is too low ($\alpha\sim10^{-5}$), the accretion rates are lower than observed. We require $\alpha\sim10^{-4}$ for the simulated discs to be consistent with the observations.}
    \label{fig:byAlpha}
\end{figure}

Figure \ref{fig:AlphaEvap} shows the mass accretion rate from our models over time, alongside observed Peter Pan disc ages and accretion rates (black points). The left-panel demonstrates that for a viscous $\alpha$ of $10^{-3}$, the accretion rate is sufficiently high that no models are compatible with the Peter Pan discs, even for extremely low evaporation rates. Only for a combination of a lower $\alpha=10^{-4}$ (low accretion) \textit{and} weak external photoevaporation ($\dot{M}_w\leq 10^{-9}\msunyr$) are our models consistent with the ages and accretion rates of observed Peter Pan discs as shown in the right-panel of fig. \ref{fig:AlphaEvap}. Note, however, that for even lower values of $\alpha$ $(10^{-5})$ the discs are indeed long lived, but with current accretion rates lower than observed, as illustrated in fig. \ref{fig:byAlpha}. Our models therefore require mass flow rates through the disc compatible with an effective $\alpha\sim10^{-4}$ to be consistent with Peter Pan discs. This is towards the lower end of the typically suggested $10^{-4}<\alpha<10^{-2}$ range \citep[e.g.][]{2009ApJ...701..260I,2010ApJ...723.1241A,2012A&A...539A...9M} though note that \cite{2016ApJ...816...25P} recently measured $\alpha \leq10^{-4}$ in HL Tau. 

\begin{figure}
    \centering
    \includegraphics[width=8.5cm]{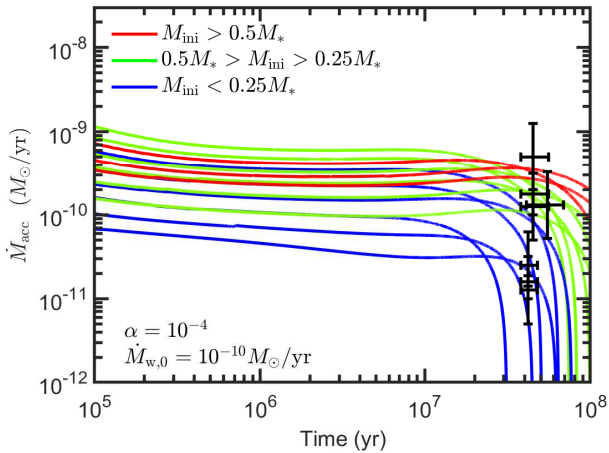}
    \vspace{-0.5cm}
    \caption{Accretion rates over time for our models with a low external photoevaporation rate and $\alpha=10^{-4}$, with selected individual models colour coded by their initial disc-to-star mass ratio. Most Peter Pan discs require $M_d > 0.25M_*$.  }
    \label{fig:byMass}
\end{figure}

\begin{figure}
    \centering
    \includegraphics[width=8.5cm]{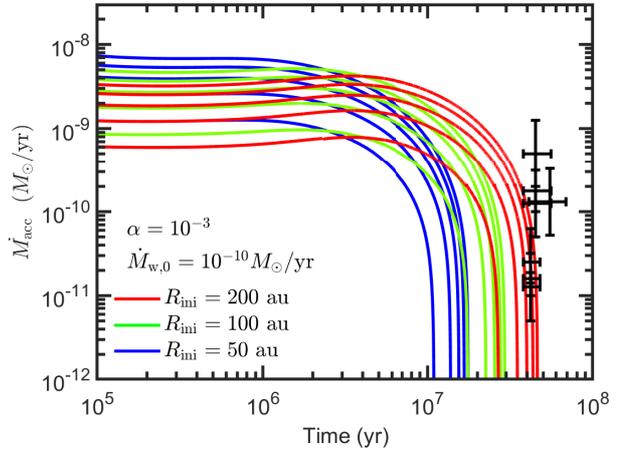}
    \vspace{-0.5cm}
    \caption{Accretion rates over time for our models with a low external photoevaporation rate and $\alpha=10^{-3}$, broken down by initial disc size. With more extended discs, disc lifetimes are longer, allowing discs with high accretions to nearly be compatible with the Peter Pan discs.}
    \label{fig:byRadius}
\end{figure}

Additionally, for those models that are consistent, we generally require higher mass discs than canonically assumed, as illustrated in fig. \ref{fig:byMass}. Most models require an initial disc-to-star mass ratio $> 0.25$ in order to both survive long enough and have high enough an accretion rate. We reiterate that high disc-to-star mass ratios can be sustained around low-mass stars \citep{2016ARA&A..54..271K, Haworth20}. For a given disc-to-star mass ratio the more extended discs also more readily yield Peter Pan discs because the viscous/accretion time-scale is longer, as illustrated in fig. \ref{fig:byRadius}. This can also allow the mass flow through the disc to be slightly larger, but even still $\alpha=10^{-3}$ still appears too strong for results to be consistent with the observations.

\vspace{-0.2cm}

\vspace{-0.2cm}
\section{Discussion}
\vspace{-0.2cm}
\subsection{Massive discs around low-mass stars?}
We find that Peter Pan discs require extremely low-UV environments over their lifetime ($<10\,G_0$), which are rare \citep{2008ApJ...675.1361F, 2020MNRAS.491..903W}. For example, approximating the solid curve of fig. 9 from \cite{2008ApJ...675.1361F} as a Gaussian, $<0.1$\,per cent of discs would experience a UV field $<10\,G_0$. This drops to 0.01\,per cent for 5$G_0$ and $\sim10^{-5}$\,per cent for 1$G_0$. Note though that this is for all Galactic star formation and that nearby low-mass star-forming regions such as Taurus are sites of such weaker UV fields. Though even in such a region a prospective Peter Pan disc would need to be on the periphery or be embedded for a significantly long time to be shielded from the stronger UV fields in the denser part to the cluster, since even a short time (e.g. $<1$Myr) of exposure to higher ($>10\,G_0$) UV environments would induce significant external photoevaporation rates, shortening the lifetime of the disc (see Figure \ref{fig:AlphaEvap}).
We also find that Peter Pan discs require low mass accretion rates onto the star, compatible with an effective $\alpha\sim10^{-4}$ consistent with observed mass accretion rates for low-mass stars \citep{Manara17} as well as suggested values for $\alpha$ \citep{2009ApJ...701..260I}.

The other requirement is that the discs are initially massive and significantly extended. Initial disc-to-star mass ratios are canonically assumed to be $<25\%$, \textit{independent of the stellar mass}; however, \cite{Haworth20} recently showed that higher disc-to-star mass ratios  are stable around low-mass (e.g. $0.1\,\msun$) stars. Our results show that these higher disc-to-star mass ratios ($>25\%$ and sometimes $>50\%$) are necessary to produce the lifetimes and accretion rates of observed Peter Pan discs, even in weak UV environments and for low $\alpha$. These discs also extend to $\sim100$--$200\au$. These higher possible disc masses may be important for explaining the bountiful mass in planets being discovered around M dwarfs. However, the actual distribution of initial disc conditions is unknown. Given that the UV fields required to produce Peter Pan discs are very rare, this raises the notion that the required high initial disc mass for low-mass stars should be relatively common, which can be quantified as the population of Peter Pan discs is increased. 

\vspace{-0.2cm}
\subsection{Why are Peter Pan discs around low-mass stars?}
The question arises as to why Peter Pan discs have only been discovered around low-mass stars. To determine whether this is a consequence of only low-mass stars being capable of producing Peter Pan discs, we re-ran our models for the case of a solar mass star. Note that in this case the highest disc-to-star mass ratio is lower, following equation \ref{eq:max_disc_mass} \citep{Haworth20}. Although we find that the lifetimes and accretion rates of the observed population of Peter Pan discs cannot be fully replicated with solar mass stars, the largest, most massive discs with low external photoevaporation of $~10^{-10}\msunyr$ and  $\alpha\sim10^{-3}$--$10^{-4}$ can survive to $\sim$50 Myr. So, in principle, we conclude that substantially older discs could be found around higher mass stars similar to the current population of Peter Pan discs. The lack of detection so far could just be a result of sampling due to there being more lower mass stars from the initial mass function. However, in the $\sim10$\,Myr Upper Sco, \cite{2018AJ....156...75E} do find a higher primordial disc fraction around lower mass stars. It is  possible that  higher mass stars typically form in the denser parts of the cluster where the radiation field and hence external photoevaporative mass-loss would be higher and reduce the disc lifetimes. Alternatively, if more efficient transport of material through the disc (higher $\alpha$) was a characteristic of discs around higher mass stars, this would similarly reduce the disc lifetime.

\vspace{-0.2cm}
\subsection{Implications for planet formation}
The disc mass required for Peter Pan discs is much larger than those typically used in planet formation models; for example, \citet{Coleman19} used disc masses between 2.7 and 8\% of the stellar mass extending out to $10\au$. Interestingly, if we calculate the gas mass within $10\au$ of the discs modelled here, we find they are equal to 0.75--7.5\% of the stellar mass, comparable to the masses used in \citet{Coleman19}. Therefore, it is plausible that the planetary systems around Peter Pan discs would be similar to those formed in \citet{Coleman19} with a few subtle differences. With the larger and more extended disc, pebble accretion would continue for longer than in \citet{Coleman19}, and as such slightly more massive terrestrial and super-Earth mass planets would be able to form. Additionally, since the Peter Pan discs are much longer lived, then planet migration would occur for longer \citep[e.g.][]{BaruteauPP6}, which would result in the planets residing close to the central star. It would be expected though that the resulting planetary systems would contain a mix of systems containing long resonant chains \citep[e.g. TRAPPIST-1;][]{GillonTrappist17}, as well as systems with planets not in resonance \citep[e.g. GJ 1061;][]{Dreizler20}.

\vspace{-0.4cm}
\section{Summary and conclusions}
\label{sec:conclusions}
We use disc evolutionary models to investigate what kind of discs can have both the longevity and currently observed accretion rates of the newly discovered long-lived Peter Pan discs. We find that they require three key ingredients: \\

1) Unusually weak UV radiation environments such that the mass-loss rate is $\leq10^{-9}\msunyr$, otherwise the disc lifetime is too short (see fig. \ref{fig:AlphaEvap}).  

2) Weak transport of material through the disc. In the $\alpha$ viscosity framework, this corresponds to $\alpha\sim10^{-4}$. A higher (canonical) accretion rate of $\alpha\sim10^{-3}$ yields too short a disc lifetime and weaker transport $\alpha\sim10^{-5}$ yields too low an accretion rate (see fig. \ref{fig:byAlpha}).

3) Even with (1) and (2), disc masses $>0.25$ times the stellar mass are typically required (see fig. \ref{fig:byMass}). \\

The constraints on external photoevaporation alone imply that Peter Pan discs are relatively rare, since they require extremely weak (and rare) UV environments of $<10$\,G$_0$ \citep{2008ApJ...675.1361F}. This corresponds to being on the periphery of low-mass star-forming regions, or perhaps more likely, deeply embedded. Given that the required environments are rare, the requirements for the mass flow through the disc and the initial disc mass may provide some insight into the initial conditions of the wider population of discs around low-mass stars. These have important implications for planet formation around low-mass stars. Higher disc-to-star mass ratios are important for setting the mass budget in protoplanetary disc, while the mass flow through the disc impacts on the disc evolution and the movement of solids within the disc.

\vspace{-0.4cm}
\section*{Acknowledgements}
We thank the anonymous referee, Cathie Clarke, and Richard Nelson for useful comments on the paper.
GALC was funded by the Leverhulme Trust through grant RPG-2018-418.
TJH was funded by a Royal Society Dorothy Hodgkin Fellowship.
This research utilised Queen Mary's Apocrita HPC facility, supported by QMUL Research-IT (http://doi.org/10.5281/zenodo.438045).

\vspace{-0.2cm}
\bibliographystyle{mnras}
\bibliography{references}{}

\vspace{-0.2cm}
\label{lastpage}
\end{document}